\documentclass[12pt]{article}%
\pdfoutput=1 

\nonstopmode

\usepackage{natbib}
\usepackage{amsmath}
\usepackage{amsbsy}
\usepackage{amscd}
\usepackage{amsopn}
\usepackage{amstext}
\usepackage{amsxtra}
\usepackage{amssymb}
\usepackage{amsthm,amsfonts}
\usepackage{graphicx}
\usepackage{epstopdf}

\usepackage{epsfig}

\def\f{\frac}

\def\CA{{\cal A}}

\def\CE{{\cal E}}
\def\CF{{\cal F}}
\def\CG{{\cal G}}

\def\CK{{\cal K}}

\def\CM{{\cal M}}
\def\CN{{\cal N}}

\def\CP{{\cal P}}

\def\CS{{\cal S}}

\def\WOR{\mbox{ WOR }}
\def\supp{\mbox{supp}}
\def\phistar{\phi_\star}

\def\BBP{\mathbb{P}}

\def\BBC{\mathbb{C}}
\def\dotstar{\cdot\! \ast}
\def\BBR{\mathbb{R}}

\def\ci{\perp\!\!\!\perp}

\def\one{\mathbf{1}}

\def\beq{\begin{eqnarray}}
\def\eeq{\end{eqnarray}}
\def\beqq{\begin{eqnarray*}}
\def\eeqq{\end{eqnarray*}}
\def\beeq{\begin{eqnarray*}}
\def\eeeq{\end{eqnarray*}}
\def\be{\begin{equation}}
\def\ee{\end{equation}}

\newtheorem{theorem}{Theorem}

\newtheorem{remark}[theorem]{Remark}

\makeindex             

\begin{document}

\title{On Watts' Cascade Model with Random Link Weights}
\author{T. R.  Hurd$^1$,  James P. Gleeson$^2$\\
\emph{ $^1$ Department of Mathematics, McMaster University, Canada }\\
 \emph{$^2$MACSI, Department of Mathematics \& Statistics,}\\\emph{ University of Limerick, Ireland}}
%
%
\maketitle


\abstract{We study an extension of D. Watts 2002  model of information cascades in social networks where edge weights are taken to be random, an innovation motivated by recent applications of cascade analysis to systemic risk in financial networks. The main result is a probabilistic analysis that characterizes the cascade in an infinite network as the fixed point of a vector-valued mapping, explicit in terms of convolution integrals that can be efficiently evaluated numerically using the fast Fourier transform algorithm. A second  result gives an approximate probabilistic analysis of cascades on ``real world networks'', finite networks based on a fixed deterministic graph. Extensive cross testing with Monte Carlo estimates shows that this approximate analysis performs surprisingly well, and provides a flexible ``microscope'' that can be used to investigate properties of information cascades in real world networks over  a wide range of model parameters.}
\bigskip



\section{Introduction} 
 The main concern of the present paper is to study a specific generalization of the well-known random cascade model introduced by Duncan \cite{Watts02} that represents the spread of an ``infectious idea'', such as a fad, a rumour, a song, a political opinion, or a new technology, on a random social network.  In Watts' model, individuals and their friendship relations are modelled as nodes and undirected links of a ``random network'', i.e. a generalization of the random graph model of \cite{ErdoReny59}. Each node is assumed to have a deterministic or random threshold $\Phi$, such that a node will ``adopt'' the new idea as soon the number of its friends already adopting this idea exceeds the threshold $\Phi$.  Watts showed under some strong assumptions that the new idea will ``percolate'' or ``cascade'' through the infinite network. More precisely, this means that the set of adopting nodes triggered by any infinitesimal fraction of seed nodes will grow to a positive fraction of the network with positive probability if and only if a certain analytical condition on the network parameters holds.  

 In the present paper, we no longer assume that the strengths of social interactions between friends (i.e. linked nodes) are equal and deterministic, but instead take edge weights to be  random variables whose distributions may be dependent on the connectedness of the nodes. This means that if $\ell$ denotes a social link between two friends who have friendship degrees $k,k'$  (that is, these friends each have in total $k$ and $k'$ friends respectively), we take the strength of their friendship to be $w_\ell$, a random variable drawn from a distribution $G_{kk'}$ that depends on $k,k'$. This new type of randomness reflects a genuine uncertainty we have in our observations of real social networks. 

The problem of random edge weights appears to be little studied in the vast literature that has built up around the Watts Cascade Model. However, the theory has moved substantially in other directions and promising new areas of application have been found  in\cite{WattDodd07}, \cite{DoddPayn09}, and\cite{DodHarDan12}. These developments are  reviewed in \cite{Newman10}. The Watts percolation theory itself has been extended to far more general random graph ensembles, for example in \cite{BoguSerr05}.

 Our motivation to study the Watts model in this way comes from a relatively new thread of economic research into ``systemic risk'' in financial networks. This term refers to financial system crises that involve domino-type cascades of defaulting or insolvent banks, triggered by an initial shock.  \cite{EiseNoe01} provided a highly stylized network picture of a  system of banks with balance sheets that distinguish external and interbank lending and borrowing. In this network, each directed edge between banks has a weight that denotes the size of an interbank loan. An initial shock to the balance sheets of the system may cause the insolvency of some banks. Insolvent bank assets are liquidated at less than their book value,  causing balance sheet shocks to their creditor banks. Such shocks may lead to secondary insolvencies: understanding the resultant ``insolvency cascade'' is the focus of systemic risk research. \cite{NieYanYorAle07}, in an influential paper, provided a random specification of this type of system, and through Monte Carlo simulations, discovered a number of non-intuitive, non-monotonic relationships between model parameters and the resultant cascade size probability distribution.  \cite{GaiKapa10} then introduced an elegant systemic risk model based on a directed graph version of Watts' setup, and by adapting Watts percolation results, were able to determine cascade sizes by analytical means. This paper was extended in several directions in \cite{AminContMinc12} and \cite{HurdGlee11}, who proved more general versions of the percolation theorems. 

This new analytical theory of systemic risk hits a major roadblock whenever a realistic specification of interbank lending is made: until now, such models have not able to handle edge weights that are random or depend deterministically on the size of banks. It is this roadblock in understanding systemic risk that we address here. To keep everything as simple as possible, we turn away from financial systems modelled by directed networks, and focus instead on information cascades in undirected networks that is the standard interpretation of Watts model.   

Our version of the Watts undirected cascade model allows for a lot of modelling flexibility, since the underlying network (the ``skeleton graph''), the adoption thresholds, the edge weights and the initial set of ``seed nodes'' are all allowed to be stochastic. One might take the view that the theory is now far from parsimonious. Making use of this additional flexibility in practice comes at a cost: statistical estimation or ``calibration'' of the myriad parameters to any specific real world network may turn out to be prohibitively difficult. It is this concern that lead us to the second main contribution of our paper: we are able to specialize the framework to any deterministic finite skeleton graph, thereby reducing complexity, whenever one accurately observes the nodes and links (as one does for example in Facebook networks, \cite{TraKelMucPor11}). The resultant picture is then of a known ``real world network'', but with randomness in its node thresholds, edge weights, and initial shocks that directly reflects our remaining uncertainties. As we will see in this paper, the Watts type cascade analysis extends naturally to this  more realistic setting.

Modelling real world networks directly in this way contrasts with prior work on real world networks that makes use of the infinite network analysis, \cite{MelHacMasMucGle11}. Typically, these papers start with network data thought of as an empirical probability distribution and as a first step, match this empirical distribution to an estimated distribution, taken from a parametric family of infinite random graph models. The cascade analysis is then applied to this estimated infinite graph model. Finally, the resultant cascade probabilities are taken as indicators of the cascade risk inherent in the original real world network. The conclusions of such an indirect approach will likely be weak, since the structure of a given real observed network is often very unlike any of the standard families of infinite random graph models.

By comparison, our probabilistic cascade analysis works directly on the empirical finite network. Whenever the analytic hypotheses are verified, or shown to have insignificant impact, we obtain measures of cascade risk that directly reflect the actual topology and structure of the real world network. We can then hope that our predicted cascade measures will accurately reflect reality. In this paper we will see whether this hope is justified by comparing our analytical formulas to Monte Carlo estimates under a range of modelling specifications.  

During this work certain subtleties concerning the dependence amongst random variables lead us to reevaluate the logical and probabilistic arguments for cascading that are typically invoked in random graph theory. In the literature, for example \cite{Newman10}, such arguments have rested heavily on the use of generating functions and underplayed the role of conditional probability. The logical framework is certainly elegant and efficient. However, we have found it preferable to use a more direct and transparent probabilistic style of argument initiated in \cite{HurdGlee11}.  It would be a demanding exercise to translate our probabilistic analysis into the more familiar generating function language! 

 Our analytical results only hold under an important and restrictive assumption that generalizes a property of the random configuration graph model of \cite{Bollobas80} that was the setting of Watts' original paper. 
Called the locally tree-like independence assumption (LTIA) in \cite{HurdGlee11}, this is the formal equivalent to the so-called ``mean-field'' assumptions often made in the literature. It holds in the infinite graph limit in the class of configuration models, but fails to hold in many other important classes of random graph models. The second technical innovation of our paper, the WOR condition, makes sense only under the LTIA assumption. It provides the key to finding the somewhat subtle unravelling of dependent random variables needed to prove Theorem \ref{thm1}.  

The flexibility of our analytical framework opens up a lot of questions for study. Space permits us to address only a few: Do we observe qualitatively new network effects such as phase transitions, due to the randomness of edge weights? Do results from the infinite graph setting accurately reflect what is observed in finite networks? When does the LTIA assumption lead to accurate results for real world networks? We hope readers will find the methods developed here compelling enough to program for themselves, and invite them to use such methods as a computer ``microscope'' to investigate further detailed properties of random networks.

%

\section{The Extended Watts Cascade Model}
The famous paper \cite{Watts02} introduces a model of a social network (network of ``friendships''): the problem is to study if and how a newly introduced technology ``percolates'' through the network. Such information cascades can be viewed as a model for the spread of ideas, rumours, a social fad, or any number of analogous phenomena. This paper has important antecedents in the literature, notably \cite{Granovetter78} and \cite{Schelling73} (see also \cite{EaslKlei10} and references therein). The setup of the original model starts with a ``skeleton graph'' on $N$ nodes labelled by $\CN=\{1,2,\dots, N\}$ (where $N$ may be infinite), that is, an undirected random graph $\CE\subset\CN\times\CN$. Each undirected link of the graph represents the existence of a friendship connection between two individuals or nodes. For the next layer of model structure, each node $v\in \CN$ is assigned a random threshold $\Phi_v\in\BBR$. At the top level, a random subset $\CA_0\in\CN$ of ``early adopter'' nodes is selected that represent those individuals that adopt the new technology before being influenced by friends.  For each realization of the above set of random variables, the cascade describing the adoption of technology by nodes proceeds through a sequence of deterministic steps. At each step, a node $v$ ``adopts''  if at least $\Phi_v$ of its neighbours were in ``adopted'' state in the previous step. The primary aim of the model is to characterize the probability of nodes eventually adopting the technology, and in particular to determine the ``phases'' in which the new technology will be taken up by a large fraction of the network. 
 
There are  many possible generalizations of the Watts model. In this paper we focus on the particular generalization where the edges $\ell$ are given random weights $w_\ell$ that describe the degree of friendship (or influence) between friends. The random nature of these weights reflects the uncertainty we have in knowing the strength of friendships in a network. 
 One is now interested in questions such as: \begin{itemize}
  \item Is there a major effect on the cascade if $w_\ell$ has fat tails (such as a Pareto distribution)?
  \item Does the percolation picture of cascades survive in this generalization?
  \item Are the cascade results computable in closed form, or must they be replaced by numerical techniques?
  \item Do random edge weights lead to interesting new phase structures as model parameters are varied?  
\end{itemize}
The full specification of the extended model consists of the following ingredients. \begin{enumerate}
  \item The skeleton graph $\CE$ is a random Bollob\'as configuration graph with a specified degree distribution $P_k=\BBP[\deg(v)=k], k\in\CK$. For simplicity, we assume that $\CK$ is a finite set of non-negative integers. We also let $z=\sum_k k P_k$ denote the mean degree. 
  \item The random variables $\Phi_v\in\BBR^+$ have cumulative distribution functions (CDFs)  $F_k(x)=\BBP[\Phi_v\le x|k_v=k]$ that depend only on the degree $k_v$. We suppose that the set of initially adopting nodes $\CA_0$ is defined by the condition $\Phi_v=0$, which occurs with probability $A^{(0)}_k=F_k(0)$ if $k_v=k$. 
    \item The random weights $w_{vv'}\in\BBR^+$ for edges $(v,v')\in\CE$ have CDFs $G_{kk'}(x)$ which depend on the degrees $k_v=k, k_{v'}=k'$. 
\item The condition for a node $v$ to have ``adopted'' at the $n$th step of the cascade, written $v\in\CA_n$,  is
 \be\label{threshold} \sum_{v'\in\CN_v\cap\CA_{n-1}} w_{v'v} \ge \Phi_v \ee
  for any $n>0$.
 \end{enumerate}
Finally, we assume a strong form of conditional independence: conditioned on the skeleton graph $\CE$, the family of random variables $\{\Phi_v,w_{vv'}\}$ for all $v,v'$ form a mutually independent collection.
 
Note that the original Watts model arises by taking $\CE$ to be a configuration random graph with general degree distribution, $w_\ell=1$ deterministically and $F_k(x)=A_k^{(0)}+(1-A_k^{(0)})F(x/k)$ for a single function $F(x)$.

%

Our aim is now to track the evolution of the cascade through the increasing sequence of sets of nodes:
\[\CA_0\to\CA_1\to\cdots\to\CA_n\to\cdots\ \]
and to determine the probabilities that a node of given degree eventually adopts:
\[A^{(\infty)}_k=\lim_{n\to\infty}\BBP[v\in\CA_n|k_v=k]\ .
\]

The mathematical analysis that follows is based on a single assumption we will make throughout this paper.
\bigskip

\noindent {\bf The ``locally tree-like'' independence assumption (LTIA):\ } Consider an extended Watts model on $\CN$ defined by a collection of random variables $(\CE, \{\Phi\}, \{w\})$.
Let $\CN_1,\CN_2\subset\CN$ be any two subsets that share exactly one node $\CN_1\cap\CN_2=\{v\}$ and let $X_1,X_2$ be any pair of random variables where  for each $i=1,2$, $X_i$ is determined by the information on $\CN_i$. Then, conditioned on information located at the node $v$, that is conditioned on $\Phi_v$ or $k_v$,  $X_1$ and $X_2$ are independent.\footnote{The following is a concise mathematical expression of this independence assumption. To any subset of nodes $\CN'\subset\CN$ we associate the sigma-algebra $\CG'$ generated by the thresholds $\Phi_v$ and degrees $k_v$ of nodes  $v\in\CN'$ and edges $(v,v')$ in $\CN'\times\CN'$. If we let $\CG_1,\CG_2,\CG_v$ denote the corresponding sigma-algebras, then $\CG_1$ and $\CG_2$ are independent sigma-algebras conditioned on $\CG_v$: that is, $ \left(\CG_1 \ci \CG_2\right)|\CG_v$.}

\bigskip

LTIA is a concise formalization of conditions that appear extensively in the literature under the name of ``mean field'' approximations. In our setting, two distinct model properties are captured by LTIA:
\begin{enumerate}
  \item The random graph $\CE$ should be tree-like in the sense that the density of triangles and higher order cycles is zero. This is true if $\CE$ is an actual finite tree (if $N$ is finite), or if it is drawn from an infinite Bollob\~as configuration model.
  \item The collection of random variables $\Phi_v,w_{vv'}$ form a mutually independent collection, conditioned on the random skeleton graph $\CE$. 
\end{enumerate}

When the LTIA condition is exactly true, the analytical results we present here are exact. In practice, in a model where LTIA is not true, our analytical results are an approximation to the true behaviour in the model. 

Certain mathematical considerations arise in the following analysis. First, one needs to know that the degree distribution of $v$ conditioned on $v$ being the end-node of an edge is not $P_k$ but 
 \[Q_k=\BBP[k_v=k|v\in\CN_{v'}\mbox{ for some $v'$}]=\f{kP_{k}}{z}\ , k\in \CK\]
 where $z=\sum_k kP_k$ is the mean degree.
 
The second consideration is to have a compact representation for the probability that $X\ge Y$ for two independent non-negative random variables. Let $\langle\cdot,\cdot\rangle$ denote the standard Hermitian inner product  on functions on $(-\infty,\infty)$:
\[\langle f,g\rangle=\int^\infty_{-\infty} \bar f(x) g(x) dx \ .\] 
Assuming that $X$ has a density $f_X(x)$ and the CDF $F_Y$ of $Y$ is continuous, and letting these functions have Fourier transforms $\CF(f_X),\CF(F_Y)$, then by the Parseval Identity
\be\label{Parseval} \BBP[X\ge Y]=\int^\infty_0\ f_X(x)F_Y(x)dx= \langle F_Y, f_X\rangle =\f1{2\pi} \langle \CF(F_Y),\CF(f_X)\rangle .
\ee
The last of these formulas will be useful in computing such probabilities when $X$ is itself a sum of independent random variables $X=X_1+X_2$ because whereas $f_X(x)$ is a convolution 
\[ f_X(x)=\int^x_0 f_{X_1}(y)f_{X_2}(x-y) dy \ ,\]
its Fourier transform is a simple product, $\CF(f_{X})=\CF(f_{X_1})\CF(f_{X_2})$.

The main result of this section is a theorem that determines the double sequence of probabilities
\[ A^{(n)}_k=\BBP[v\in\CA_n|k_v=k]\]
beginning with the initial vector $A^{(0)}_k, k\in\CK$. The theorem is inductive  on the steps $n=0,1,2,\dots$ of the cascade. To motivate the result, we consider the defining property \eqref{threshold} of the set $\CA_n$ and note this implies
\be\label{induction} A^{(n)}_k=\BBP[ \Phi_v \le \sum_{v'\in\CN_v\cap\CA_{n-1}} w_{v'v} |k_v=k] \ .\ee
The subtle point is that the random variables $\Phi_v$, $\CA_{n-1}$ and $w_{vv'}$ occurring  in this expression have a complicated dependence, and the probability cannot be evaluated directly using \eqref{Parseval}. The key to unravelling this dependence is a notion that takes care of the causal relations in the order of adopting nodes. We say that a node property $\CP$ holds for a node $v$ ``without regarding'' a neighbouring node $v'\in\CN_v$, and write ``$\CP$ holds for $v$ WOR $v'$'', if this property of $v$ holds in the subgraph of $\CE$ obtained by deleting the edge to $v'$ and the entire connected subgraph rooted on this edge. It is important to note that the concept  ``$\CP$ holds for $v$ WOR $v'$'' makes sense only under the LTIA assumption: otherwise with positive probability the deleted subgraph will include $v$ itself via a cycle in the network. 

We will show that \eqref{induction} is expressible in terms of intermediate WOR probabilities defined by
\be C^{(n)}_k=\BBP[v\in\CA_n \WOR v'|k_v=k,v'\in\CN_v]\ee
where $v,v'$ are random nodes selected from $\CN$. At the outset,
 \be C^{(0)}_k=\BBP[\Phi_v=0|k_v=k,v'\in\CN_v]= A^{(0)}_k\ .
 \ee

We begin by noting that for $n>0$, $v\in\CA_n$ is true if and only if 
\[\Phi_v \le \sum_{v''\in\CN_v} \tilde w^{(n-1)}_{v'',v} \]
where 
\[\tilde w^{(n-1)}_{v'',v}=  w_{vv''}\one(v''\in\CA_{n-1} \WOR v)\ .\]
Next we observe that under the condition $k_v=k$, LTIA implies that  the $k$ random variables $\tilde w^{(n-1)}_{v'',v}$ for $v''\in\CN_v$ are independent. They are also identically distributed with CDF given by
\beqq
\BBP[\tilde w^{(n-1)}_{v'',v}\le x|k_v=k,  v''\in\CN_v]&=& \sum_{k''} \BBP[\tilde w^{(n-1)}_{v'',v}\le x|k_v=k, k_{v''}=k'', v''\in\CN_v] Q_{k''}\\
&=&  \sum_{k''} \left[(1-C^{(n-1)}_{k''}) +C^{(n-1)}_{k''}G_{kk''}(x)\right]Q_{k''}\ .
\eeqq
In the above step, we have used LTIA again: conditioned on $k_v=k, k_{v''}=k''$, the random variables $w_{vv''}$ and $ \one(v''\in\CA_{n-1} \WOR v)$ are independent.
Hence it follows that 
\beq A^{(n)}_k&=&\BBP\Bigl[\sum_{v''\in\CN_v} w_{vv''}\one(v''\in\CA_{n-1} \WOR v) \ge \Phi_v|k_v=k\Bigr]\nonumber\\
&=&\f1{2\pi} \langle \CF(F_k),(\CF(g^{(n-1)}_k))^k\rangle \label{Aequation} 
\eeq
where $\CF(g^{(n-1)}_k)$ is the Fourier transform of 
\be\label{intermedFT}g^{(n-1)}_k(x)=\sum_{k'}\left[(1-C^{(n-1)}_{k'})\delta_0(x) + C^{(n-1)}_{k'}g_{kk'}(x)\right]Q_{k'}\ee
and $\delta_0$ denotes the Dirac delta function centred at $0$.

We now suppose the vector $\{C^{(n-1)}_k, k\in\CK\}$  is known for some $n>0$, and seek a formula for the vector $C^{(n)}_k$. The logic is the same as in the previous paragraph, with one difference. Given $k_v=k, v'\in\CN_v$,  the sum over $v''$ is over the set $\CN_v\setminus v'$ instead of the set $\CN_v$. It follows that 
 \be\label{Crecursion} C^{(n)}_k=\f1{2\pi} \langle \CF(F_k),(\CF(g^{(n-1)}_k))^{k-1}\rangle 
\ee
with $g^{(n-1)}_k$ given again by \eqref{intermedFT}.

The above arguments prove our first main result:
\begin{theorem}\label{thm1}
Consider an extended Watts model on $\CN$ defined by a collection of random variables $(\CE, \{\Phi\}, \{w\})$ that satisfies LTIA. Then: \begin{enumerate}
  \item The probabilities $C^{(n)}_k=\BBP[v\in\CA_n \WOR v'|k_v=k,v'\in\CN_v]$ satisfy the recursive formula 
\be \label{Crecursion2} C^{(n)}_k=\f1{2\pi} \langle \CF(F_k),(\CF(g^{(n-1)}_k))^{k-1}\rangle 
\ee
with $F_k(x)=\BBP[\Phi_v\le x|k_v=k]$, $g_k^{(n-1)}(x)$ given by \eqref{intermedFT}. The initial values are $C^{(0)}_k=A^{(0)}_k=\BBP[\Phi_v=0|k_v=k]$. 
  \item The probabilities $A^{(n)}_k=\BBP[v\in\CA_n|k_v=k]$ are given in terms of $C^{(n-1)}_k$ by
  \be A^{(n)}_k = \f1{2\pi} \langle \CF(F_k),(\CF(g^{(n-1)}_k))^k\rangle .\label{Aequation2} \ee
  
  \item The probability that a node of degree $k$ eventually adopts, $A^{(\infty)}_k$, is given by \eqref{Aequation2} 
 in terms of a vector $C^{(\infty)}=\{C^{(\infty)}_k\}_{k\in\CK}$ that satisfies the $\mathbb{R}^K$ valued fixed point equation defined by the recursion given in part (1). The existence (but not uniqueness) of such a fixed point is guaranteed by the Tarski-Knaster Fixed Point Theorem. 
\end{enumerate}
\end{theorem}

\begin{remark}  It is interesting that nearest-neighbour correlations in this model are determined in terms of the $C$ and $A$ probabilities. In fact, under LTIA, the joint probability of eventual non-adoption of a nearest neighbour node pair $v,v'$ with $k_v=k,k_{v'}=k'$ is simply:
\beqq\BBP[v,v'\notin\CA_\infty|k_v=k,k_{v'}=k']&=&\BBP[v\notin\CA_\infty\WOR v',v'\notin\CA_\infty\WOR v|k_v=k,k_{v'}=k'] \\&=& [1-C_{k'}^{(\infty)}][1-C_{k}^{(\infty)}]\ .\label{corr}
\eeqq
A result of this type is found in \cite{GleeMeln09}. 
\end{remark}

It is well-known  in network science, see e.g. \cite{Newman10}[Chapter 13.11], that the frequency of global cascades in infinite random graphs is related to  the fractional size of the so-called in-component associated to the giant vulnerable cluster. In the present context, a vulnerable cluster has the meaning of a connected subgraph of the network consisting of ``vulnerable'' directed edges, where a vulnerable directed edge is an edge whose weight is sufficient to exceed the adoption threshold of its forward node. We define: \begin{itemize}
\item $\CE_V\subset\CE$, the set of vulnerable directed edges;
  \item $\CE_s$, the giant strongly connected set of vulnerable edges (the ``giant vulnerable cluster'');   \item $\CE_i$, the ``in-component'' of the giant vulnerable cluster,  i.e. the set of vulnerable edges that are connected to $\CE_s$ by a directed path of vulnerable edges;
  \item $1-b_{k'}:=\BBP[(v',v)\in \CE_i|k_{v'}=k']$, a conditional probability of an edge being in  $\CE_i$;
  \item $a_{k'k}=\BBP[ \Phi_v\le w_{v'v}|k_{v'}=k',k_v=k]$, the conditional probability of a node being vulnerable.
\end{itemize}
Now note that  $(v',v)\in \CE_i^c$ (i.e. the complement of $\CE_i$) means either $\Phi_v> w_{v'v}$ or $\Phi_v\le w_{v'v}$ and all the $k_v-1$ ``downstream'' directed edges $(v,v'')$ are in the set $\CE_i^c$. Thus, invoking the LTIA, one determines that 
\beqq b_{k'} &=&\sum_k \BBP[k_v=k|k_{v'}=k', v\in \CN^-_{v'}]\BBP[(v',v)\in \CE_i|k_{v'}=k',k_v=k]\\
&=&\sum_k\frac{kP_k}{z}\left[(1-a_{k'k})+a_{k'k}b_k^{k-1}\right]:= h_{k'}({\bf b}) 
\eeqq
In other words, the vector ${\bf b}=\{b_k\}_{k\in\CK}$  satisfies the fixed point equation ${\bf b}={\bf h}({\bf b})$ where
\be h_{k'}({\bf b})= \sum_{k} \frac{kP_{k}}{z}\left[(1-a_{k'k})+a_{k'k}b_{k}^{k-1}\right], \quad k''\in\CK\ .
\ee

The cascade condition is the condition on a random network that determines whether or not it has a giant vulnerable cluster,  or equivalently, whether a randomly selected initial adopting node has a positive probability of triggering a cascade that reaches a positive fraction of the infinite network. As shown in a similar setting in \cite{DodHarPay11}, \cite{AminContMinc12} and \cite{HurdGlee11}, this boils down to a condition on the spectral radius $\|D\|$ of the derivative matrix $D=(D_{k'k})_{k'k\in\CK}$. Here $D_{k'k}=\partial h_{k'}/\partial b_{k }$ is evaluated at the trivial fixed point ${\bf b}=[1,1,\dots, 1]$, and is given by an explicit formula:
\be\label{Dmatrix}
D_{kk'}=\frac{ a_{k'k}k(k-1)P_{k}}{z}\ .
\ee 
Recall that the spectral radius of a square matrix is the magnitude of its largest eigenvalue.\\
{\bf Cascade Condition:\ } If $\|D\|>1$, then with positive probability, a single random seed will generate a cascade that reaches a positive fraction of the infinite network. If $\|D\|<1$, then with probability $1$, the cascade generated by a single random seed will  be a zero fraction of the infinite network. 

For a global cascade to arise from a random single seed, it is sufficient for the seed $v$ to trigger at least one edge $(v,v')\in\CE_i$. Therefore, by conditioning on the degree of $v$ one finds that the frequency of global cascades is bounded below:
\[ f\ge \sum_{k>0}[1- b_k^k ]P_k\ .\]

\section{Discrete Probability Distributions}

The structure of \eqref{Crecursion2} and similar equations is problematic from the point of view of numerical approximations. Numerical evaluation of the implied integrals leads to truncation errors and discretization errors, both of which will be difficult to handle in our setting. In this section, we analyze the case where the random variables $\{\Phi_v,w_\ell\}$ all take values in a specific finite discrete set $\CM=\{0,1,\dots,(M-1)\}$ with a large value $M$. In such a situation, the convolutions  in  \eqref{Aequation} can be performed exactly and efficiently by use of the discrete Fast Fourier Transform (FFT). 

Let $X,Y$ be two independent random variables with probability mass functions (PMF) $p_X,p_Y$ taking values on the non-negative integers $\{0,1,2,\dots\}$. Then the random variable $X+Y$ also takes values on this set and has the probability mass function (PMF)
$ p_{X+Y}=p_X*p_Y$ where the convolution of two functions $f,g$ is defined to be
\be\label{convolution} (f*g)(n) = \sum_{m=0}^n f(m) g(n-m) \ .\ee
Note that $p_{X+Y}$ will not necessarily have support on the finite set $\CM$ if  $p_X,p_Y$ have support on $\CM$. This discrepancy leads to the difficulty called ``aliasing'' in the next section. 

\subsection{The Discrete Fourier Transform}
In this section we consider space  $\BBC^M$ of $\BBC$-valued functions on $\CM=\{0,1,\dots,M-1\}$. The discrete Fourier transform, or fast Fourier transform (FFT), is the linear mapping $\CF:a=[a_0,\dots,a_{M-1}]\in\BBC^M\to \hat a=\CF(a)\in \BBC^M$ defined by
\[ \hat a_k=\sum_{l\in\CM} \zeta_{kl} a_l\ , k\in\CM\ . \]
where the coefficient matrix $Z=(\zeta_{kl})$ has entries $\zeta_{kl}=e^{-2\pi i kl/M}$. 

It is easy to prove that its inverse, the ``inverse FFT'' (IFFT), is given by the map $a\to\tilde a=\CG(a)$ where 
\[\tilde a_k=\frac1{M}\sum_{l\in\CM} \bar\zeta_{kl} a_l\ , k\in\CM\ . \]
If we let $\bar a$ denote the complex conjugate of $a$, we can define the Hermitian inner product between 
\[  \langle a,b\rangle:=\sum_{m\in\CM} \bar a_m b_m\ .
\]
We also define the convolution product of two vectors:
\[ (a*b)(n)= \sum_{m\in\CM}a(m)\ b(n-m\mbox{ modulo $M$} ), \quad n\in\CM \ .\]

Now we note the following easy-to-prove identities which hold for all $a,b\in \BBC^M$:\begin{enumerate}
 \item Inverse mappings:
 \[ a=\CG(\CF(a))=\CF(\CG(a))\ ;\] 
 \item Conjugation:   \[ \overline{\CG(a)}=\frac1{M}\CF(\bar a)\ ;\]
 \item Parseval Identity:
\[ \langle a,b\rangle=M\langle \tilde a,\tilde b\rangle=\frac1{M}\langle \hat a,\hat b\rangle\ ;\]
  \item  Convolution Identities:  \[ \tilde a\dotstar\tilde b=\widetilde{(a*b)}\ ; \quad \hat a\dotstar\hat b=\widehat{(a*b)}\]
  where $\cdot*$ denotes the component-wise product. 
\end{enumerate}

We note that if the support of the sum of two $\CM$-valued random variables $X, Y$ is itself in $\CM$, that is, if $\supp(X+Y)=\{n|\exists m\in\supp(X) \mbox{s.t. } n-m\in\supp(Y)\}\in\CM$, then $p_{X+Y}$, the PMF of $X+Y$, is given by $p_X*p_Y$. On the other hand, if there are $m\in\supp(X), n\in\supp(Y)$ such that $m+n\ge M$, then $p_{X+Y}-p_X*p_Y$ is not zero: such a difference is called an ``aliasing error''. In what follows we assume that all aliasing errors can be made zero by choosing $M$ sufficiently large but finite.

We now consider the  probability 
\[ P=\BBP[Y\le \sum_{i=1}^k X_i]\ . \]
for independent random variables $Y,X_1,X_2,\dots,X_k$ where $X_i$ are  distributed with PMF $g_i(m)$ and $Y$ has PMF $f(m)$. We suppose that $f$ has support on $\CM$ while each $g_i$ has support on $\{0,1,\dots, \lfloor M-1/k\rfloor\}$. We also define the CDF $F(n)=\sum_{m=0}^nf(m)=\BBP(Y\le n)$ for $n\in\CM$.  Then using the above identities, we are lead to an efficient means to compute $P$ for large $k$ involving matrix operations and the FFT:
\beq&=&\BBP[Y\le \sum_{i=1}^k X_i]=\sum_{m\in\CM}(g_1*\dots*g_k)(m)\ \sum_{n=0}^m f(n)=\sum_{m\in \CM} (g_1*\dots*g_k)(m)F(m)\nonumber\\
&=&\langle F,  g_1*\dots*g_k\rangle=\f1{M}\langle \hat F, \widehat{(g_1*\dots*g_k)}\rangle=\f1{M}( \hat F)'\ast( \hat g_1\dotstar\dots\dotstar\hat g_k )\ .\label{FFTidentity}
\eeq
Here  in the final expression, $A'$ denotes the conjugate transpose of a matrix $A$, $\ast$ denotes matrix multiplication between a size $[1,M]$ matrix and a size $[M,1]$ matrix, and $\dotstar$ denotes component-wise multiplication of vectors and matrices. 

\begin{remark}\label{aliasing}  There is no aliasing problem and the identity \eqref{FFTidentity} is true if $g_1*\dots*g_k$ has support on $\CM$. We formalise this requirement as the {\rm Aliasing Assumption}.  \end{remark}

\begin{remark}  The overall computation time for any numerical implementation of the cascade formulas in Theorem \ref{thm1} will be dominated by computing convolutions such as those in \eqref{Aequation2}. In particular, one can check that negligible total time will be taken in computing FFTs (each of which take of the order of $M\ln M$ additions and multiplications). One can compare the efficiency of our recommended FFT approach to the direct approach by considering a single evaluation of the convolution: to compute $f*g$ using \eqref{convolution} for $M$-vectors $f,g$ requires $M^2$ additions and multiplications, whereas to compute $\hat f\dotstar\hat g$  requires only $M$ multiplications. All else being equal, we can expect a speedup by a factor of $O(M)$ when the FFT method is used instead of the direct method. Since in our implementations we often have $M\gtrsim 2^{10}$, this is a huge improvement.  Of course, this factor is too optimistic: it will be worsened considerably when one takes care of the aliasing problem. \end{remark}

%

\section{Fixed Random Graphs}
We now consider how the formalism of Section 2 can be applied in an approximate fashion in the important special class of problems when the skeleton graph is actually known (i.e. deterministic) and finite while the thresholds and edge weights are random. This will allow us to address the question of systemic risk in tractable models of real observed financial networks.

Let $A=(A_{vv'}), v,v'\in\CN$ be the symmetric adjacency matrix of the fixed undirected graph $\CE$. We number the nodes in $\CN$ by $v=1,2,\dots, N$ and the links in $\CE$ by $\ell=1,2,\dots, L$ where $L=\sum_{1\le v<v'\le N} A_{vv'}$. 
The discrete threshold and edge weight random variables $\Phi_v, w_\ell$ are assumed to have support on the set $\CM=\{0, 1, \dots, M-1\}$. The probability that node $v$ is an early adopter is $A^{(0)}_v=\BBP[\Phi_v=0]=\BBP[v\in\CA_0]$.  The CDF of $\Phi_v$ is an $M$-vector $F_v(m)=\BBP[\Phi_v\le m]$ for $m\in\CM$, and we let $\hat F_v=\CF(F_v)$ denote its discrete Fourier transform (FFT). Similarly, $\hat g_{\ell}(m), m\in\CM$ denotes the FFT of the $M$-vector $g_\ell(m)=\BBP[w_{\ell}=m]$, the probability mass function of $w_\ell$. To ensure LTIA, it is necessary but not sufficient that $\{\Phi_v, w_\ell\}$ is a collection of mutually independent random variables. The LTIA is exact only when $\CE$ is a tree graph or if the edge weights and thresholds are deterministic. 

The aim of this section is to use LTIA to derive an approximate formula for the marginal likelihoods $A^{(\infty)}_v:=\BBP[v \mbox{ eventually adopts}]$.  In the general case, we expect that the approximation that takes the dependence between different branches of the random graph to be negligible should improve as $N$ gets large, all else being equal. We will find that the dependence of the accuracy on other parameters is not easy to understand, but that overall assuming LTIA leads to surprisingly good performance. 

The analysis will determine the required probabilities $A^{(n)}_v=\BBP[v\in\CA_n]$ via the intermediate probabilities 
\[C^{(n)}_{v,v'}=\BBP[v\in\CA_n \WOR v']\]
that will be computed for each of the $2L$ directed links $\ell=(v,v')$ and $n=1,2,\dots$. At the outset, $C^{(0)}_{v,v'}=A^{(0)}_v$. 

The analysis for a finite real graph exactly parallels the argument presented in Section 2.  First we note that $v\in\CA_n$ for $n>0$ means that
  \[\sum_{v''\in\CN_v} \tilde w_{v'',v} \ge \Phi_v\]
  where 
  \[\tilde w_{v'',v} :=w_{vv''} \one(v''\in\CA_{n-1}\WOR v)\ .\]
By LTIA, we assume that $w_{vv''} $ and $\one(v''\in\CA_{n-1}\WOR v)$  are independent for all $v''\in\CN_v$. Therefore, the sum of the $\tilde w$ over $v''$ is a sum of independent random variables, each with moment generating function (MGF) \footnote{Note: we have used the fact that the MGF of the random variable $0$ is the $M$-vector $[1,1,\dots,1]$.}
\be\label{intermedFTRW}\hat g^{(n-1)}_{v'',v}(m)=(1-C^{(n-1)}_{v'',v}) + C^{(n-1)}_{v'',v}\hat g_{v'',v}(m), m\in\CM\ .\ee
From these facts, and \eqref{FFTidentity}, and making use of the Aliasing Assumption (Remark \ref{aliasing}), we deduce the formula for $A^{(n)}_v$:
\be\label{AformulaRW} A^{(n)}_v=\BBP[\sum_{v''\in\CN_v} \tilde w_{v'',v} \ge \Phi_v]=\Bigl\langle\hat F_v,\prod_{v''\in\CN_v}\hat g^{(n-1)}_{v'',v}\Bigr\rangle\ .
\ee

When it comes to computing $C^{(n)}_{v,v'}$ in terms of the set $\{C^{(n-1)}\}$, we may repeat this argument. The only change is in the sum over $v''$: it now runs over $\CN_v\setminus\{v'\}$ not $\CN_v$. It follows that
\be\label{CformulaRW} C^{(n)}_{v,v'}=\Bigl\langle\hat F_v,\prod_{v''\in\CN_v\setminus\{v'\}}\hat g^{(n-1)}_{v'',v}\Bigl\rangle\ .
\ee

The above argument rests on two assumptions, and leads directly to the main result concerning real world graphs: 

\begin{theorem}\label{thm2}
Consider an extended Watts model on $\CN$ on a fixed finite graph $\CE$ with $N$ nodes and $L$ edges. Let the thresholds and edge weights be a collection of independent random variables $\{\Phi_v, w_\ell\}$ supported on $\CM=\{0,1,\dots, M-1\}$. Assuming LTIA and the Aliasing Assumption (Remark \ref{aliasing}), the following formulas hold: \begin{enumerate}
  \item For each $n>0$ and directed edge $v,v'$, the probabilities $C^{(n)}_{v,v'}=\BBP[v\in\CA_n \WOR v']$ satisfy the recursive formula 
\[ C^{(n)}_{v,v'}=\Bigl\langle\hat F_v,\prod_{v''\in\CN_v\setminus\{v'\}}\hat g^{(n-1)}_{v'',v}\Bigl\rangle\ 
\] 
with $F_v(m)=\BBP[\Phi_v\le m]$ and $\hat g_{v'',v}^{(n-1)}(x)$ given by \eqref{intermedFTRW}. The initial values are $C^{(0)}_{v,v'}=A^{(0}_v=\BBP[\Phi_v=0]$. 
  \item The probabilities $A^{(n)}_v=\BBP[v\in\CA_n]$ are given in terms of the collection $C^{(n-1)}_{\ell}$ by \eqref{AformulaRW}.
  \item The probability that a node $v$ eventually adopts, $A^{(\infty)}_v$, is given by \eqref{AformulaRW} in terms of the $2L$-vector $C^{(\infty)}=\{C^{(\infty)}_{v,v'}\}$ that satisfies the $\CM^{2L}$-valued fixed point equation defined by the recursion given in part (1). The existence (but not uniqueness) of such a fixed point is guaranteed by the Tarski-Knaster Fixed Point Theorem. 
\end{enumerate}
\end{theorem}

\begin{remark} {\bf Caveats:} In practice, we can arrange for the Aliasing Assumption  to be approximately true by choosing $M$  large enough that the probabilities\\ $\BBP[\sum_{v''\in\CN_v} \tilde w_{v'',v}\ge M]$ are kept sufficiently small: aliasing errors of this order of smallness can thus be expected to appear in any implementation of our formulas. However, we cannot generally arrange for LTIA to be even approximately true. The outstanding difficulty in understanding the true nature of the adoption cascade will therefore be in understanding the biases and errors that relate the LTIA formulas to the ``true values'' that can only be estimated by slow Monte Carlo simulations of the real world network. 
\end{remark}

The percolation argument of Section 3 has a straightforward analogue for real world graphs that satisfy the LTIA approximately. As before we define: \begin{itemize}
  \item vulnerable directed edges $\ell=(v',v)$ by the condition $\Phi_v\le W_{v"v}$, $\CE_V$ to be the set of such edges, and $a_\ell=\BBP[\ell\in\CE_V]$;
  \item $\CE_s$, a strongly connected cluster of vulnerable edges;
  \item $\CE_i$, the ``in-component'' of $\CE_s$;
  \item $1-b_\ell=\BBP[\ell\in\CE_i]\ .$
\end{itemize}
Treating LTIA now as an approximation, one can then show that 
\[ b_{v'v}\sim a_{v'v}\prod_{v"\in\CN_v\setminus v'}b_{vv"} +(1-a_{v'v})\ .\]
That is, the sequence ${\bf b}=\{b_{v'v}\}$ is an approximate solution of a fixed point equation ${\bf b}={\bf h}({\bf b})$. Direct differentiation at the trivial fixed point ${\bf b}=[1,1,\dots,1]$ yields the result 
\[ D_{\ell,\ell'}=\left\{\begin{array}{ll }
a_\ell      &  \mbox{if $\ell'\in\CE_\ell\setminus\ell$}  \\
  0    &\mbox{else}   
\end{array}\right.
\]
When the spectral radius of $D$ is greater than $1$ we can expect to see large cascades from the initial random seed with frequency approximately 
\[f\ge \sum_{v\in\CN}\left[1-\prod_{v'\in\CN_v} b_{vv'}\right]A^{(0)}_v\ .\]

\section{Numerical Investigations}
In this section,  we report results of some numerical experiments we have made testing the above analysis. We divided our experiments into those that addressed the infinite random graphs of Section 2, and those that addressed real world graphs as described in Section 4. In our numerical implementations, all random variables are taken to be either deterministic or discretized lognormals. The lattice spacing $\delta x$ and lattice size $M$ were always chosen carefully  to reduce any bias between the Monte Carlo and  analytic implementations.

\subsection{Experiment 1: Infinite Graphs} For simplicity, we considered a Poisson random graph model as used in the experiments of  \cite{Watts02}, with Poisson degree distributions parametrized by $z\in[0,10]$ and threshold CDFs 
 \[ F_k(x)=A^{(0)}_k+(1-A^{(0)}_k)\one(x\ge \phistar k_v)\]  with $\phistar=0.18$ and $A^{(0)}_k=10^{-4}$. In Experiment 1A, the only change was to allow the weight of an edge to depend deterministically on the degrees of its nodes, i.e. its ``type'', while keeping the average edge weight to be $1$. We treated the original specification $w_\ell=1$ as the benchmark case, and two additional cases where the weight $\bar w=\bar{w}(k,k',\alpha)$ of an edge of type $kk'$ is proportional to $(kk')^{-\alpha}$ for some $\alpha$. The first alternative with $\alpha=-0.5$ corresponds to a social network in which the  strength of edges increases as the connectivities $k, k'$ increase. The second alternative with $\alpha=0.5$ corresponds to a social network in which the edge strength of nodes decreases as the connectivities $k,k'$ increases. The case $\alpha=0$ recovers the original benchmark case.

\begin{figure}[ht]\vspace{-3.in}
\centering\hspace{-1.in}
 \includegraphics[scale=1.0]%
{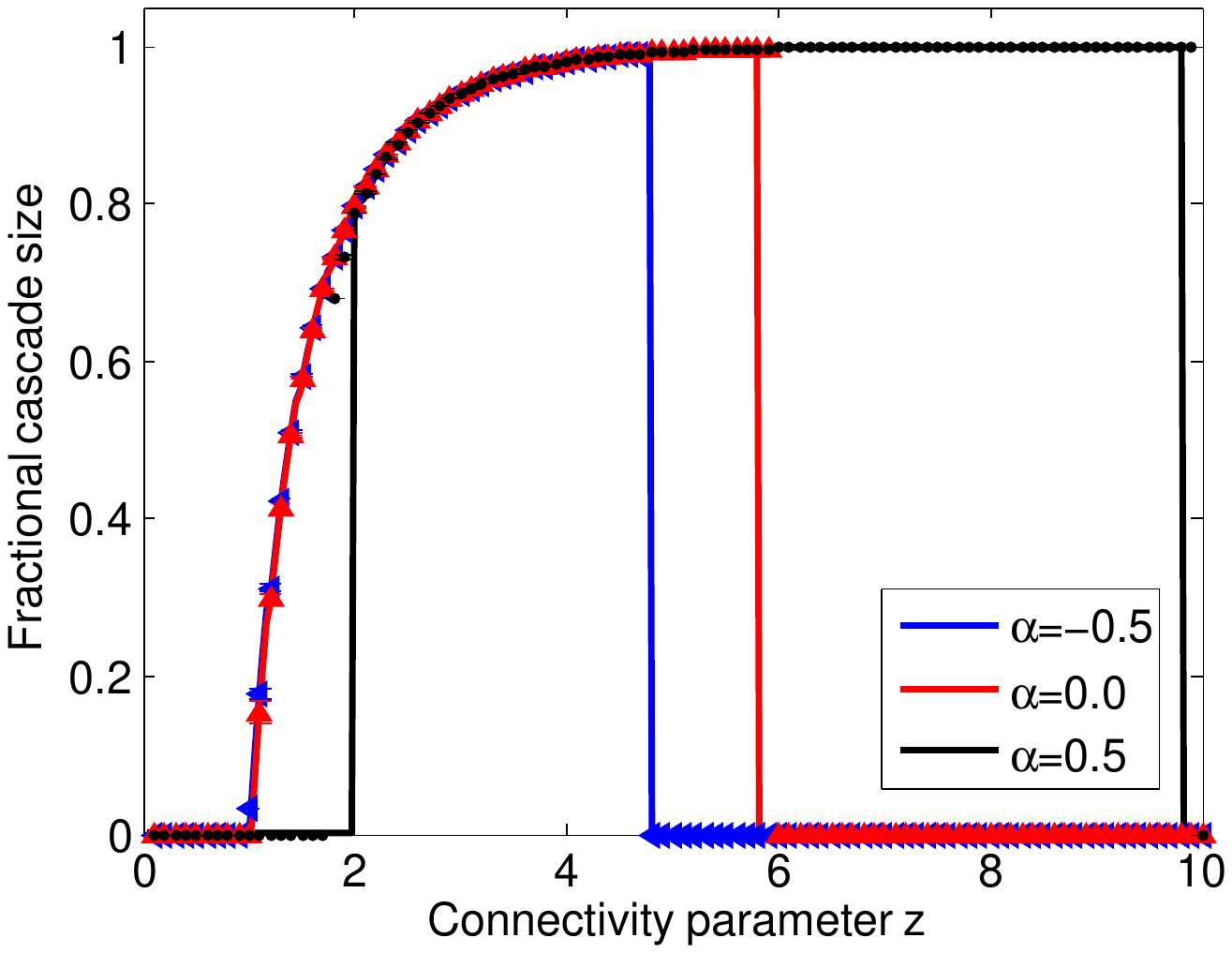}\vspace{-3.in}
\caption{Experiment 1A: This figure shows how the size of global cascades on infinite networks is influenced by nonconstant but deterministic edge weights. The benchmark case with $\alpha=0$, shown by the red triangles, is the case of constant edge weights. The cascade size remains similar in the case with $\alpha=-0.5$ (blue triangles). In both cases, the lower critical value of $z$ is approximately $1$.  On the other hand, when  $\alpha=0.5$ (black pluses), the lower cascade transition is shifted up to $z\sim 1.5$, and furthermore, the upper critical value is shifted beyond $10$ (in fact it occurs around $z=14$). } \label{fig1A}
\end{figure}

Figure \ref{fig1A} shows the results. First we note that for the default case $\alpha=0$, the curve replicates the results shown in Figure 2b of \cite{Watts02}.  The case with $\alpha<0$ does not differ dramatically from the constant weight case. Both show a lower cascade threshold very near to $z=1$.

In contrast, in the case with $\alpha>0$, where highly connected nodes have relatively less influence on their friends, we see clearly that the cascade window is shifted to higher values of $z$.  The upward shift in the lower cascade threshold away from $z=1$ is somewhat surprising, and begs an intuitive explanation. The expected transition near $z=1$ for $\alpha\le 0$ coincides almost exactly with the  percolation point at $z=1$ for Erd\"os-Renyi (ER) random graphs because, as one can check, for these values of $\alpha$ and  $z\sim1$ the vulnerable edge condition for $\CS_V$ is satisfied with high probability for all nodes. By contrast, when $\alpha=0.5$ and $z\sim 1$, there is a significant positive probability for non-vulnerable edges of low degree, and consequently a significant upward shift in the transition point. 

\begin{figure}[ht]\vspace{-3.in}
\hspace{-1.5in}
\centering \includegraphics[scale=1.0]%
{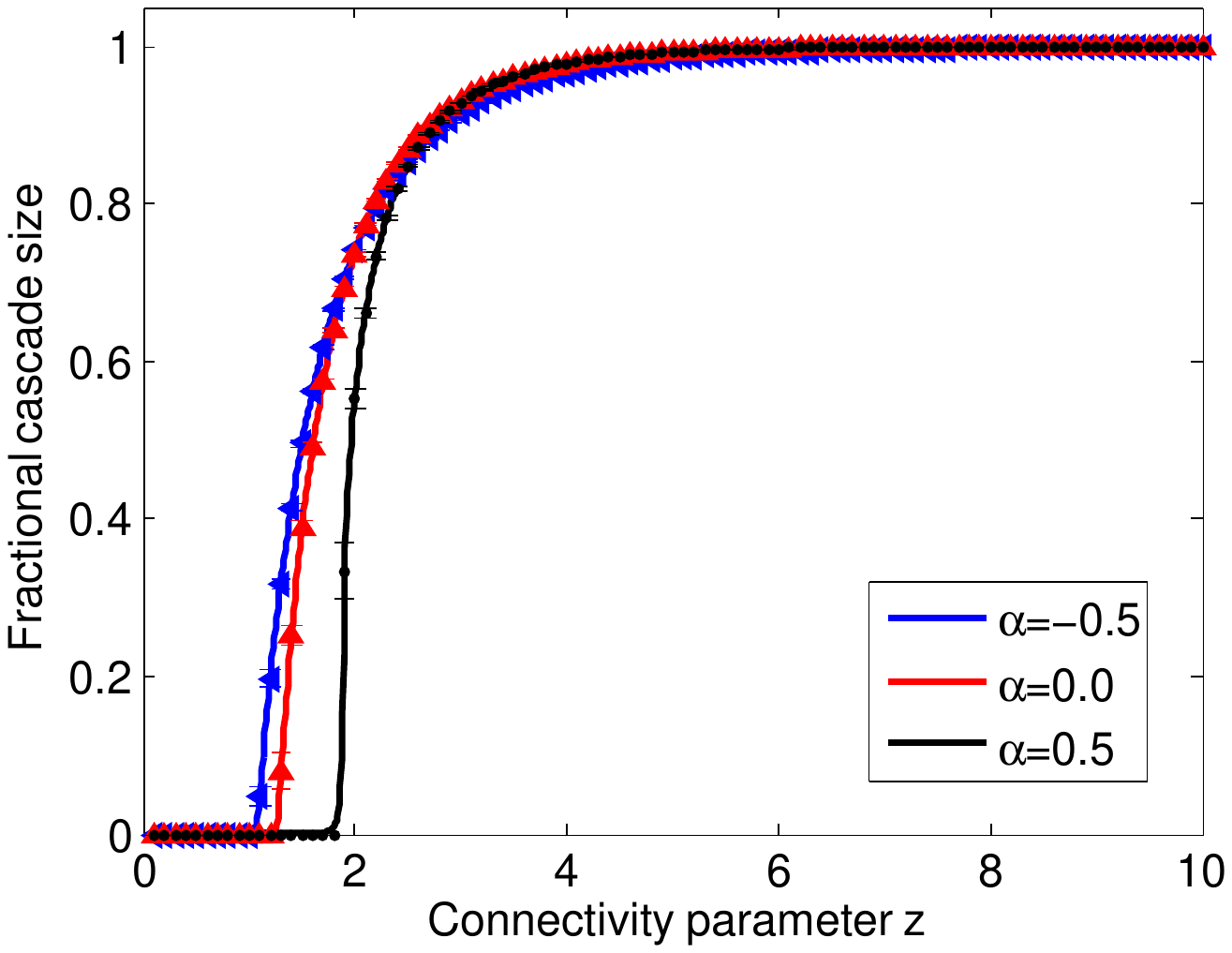}\vspace{-3.in}
\caption{Experiment 1B: This figure shows how the size of global cascades on infinite networks is influenced by lognormally distributed edge weights, with means consistent with the cases of Figure \ref{fig1A}. Compared to the deterministic cases, both the upper and lower critical values have been shifted to higher $z$ values. All three curves, if extended beyond $z=10$ exhibit upper critical values beyond $z=14$.} \label{fig1B}
\end{figure}

In Experiment 1B, we explored the effect of additional randomness, by taking the random variables $w_\ell$ to be lognormally distributed, with means $\bar w=\bar{w}(k,k',\alpha)$ matching Experiment 1A and standard deviation $0.5\times \bar w$, whenever the type of $\ell$ is $(k,k')$. Surprisingly, the effect of randomness is counter to our expectation that the critical cascade region would shift to lower $z$: as Figure \ref{fig1B} shows, the opposite behaviour occurs.

In both Experiments 1A and 1B we also compared the $N=\infty$ analytical results to large scale Monte Carlo estimates computed for $N=10^5$,  Figures \ref{fig1A} and \ref{fig1B} show the expected excellent agreement.

\begin{figure}[ht]
\centering \includegraphics[scale=.6]%
{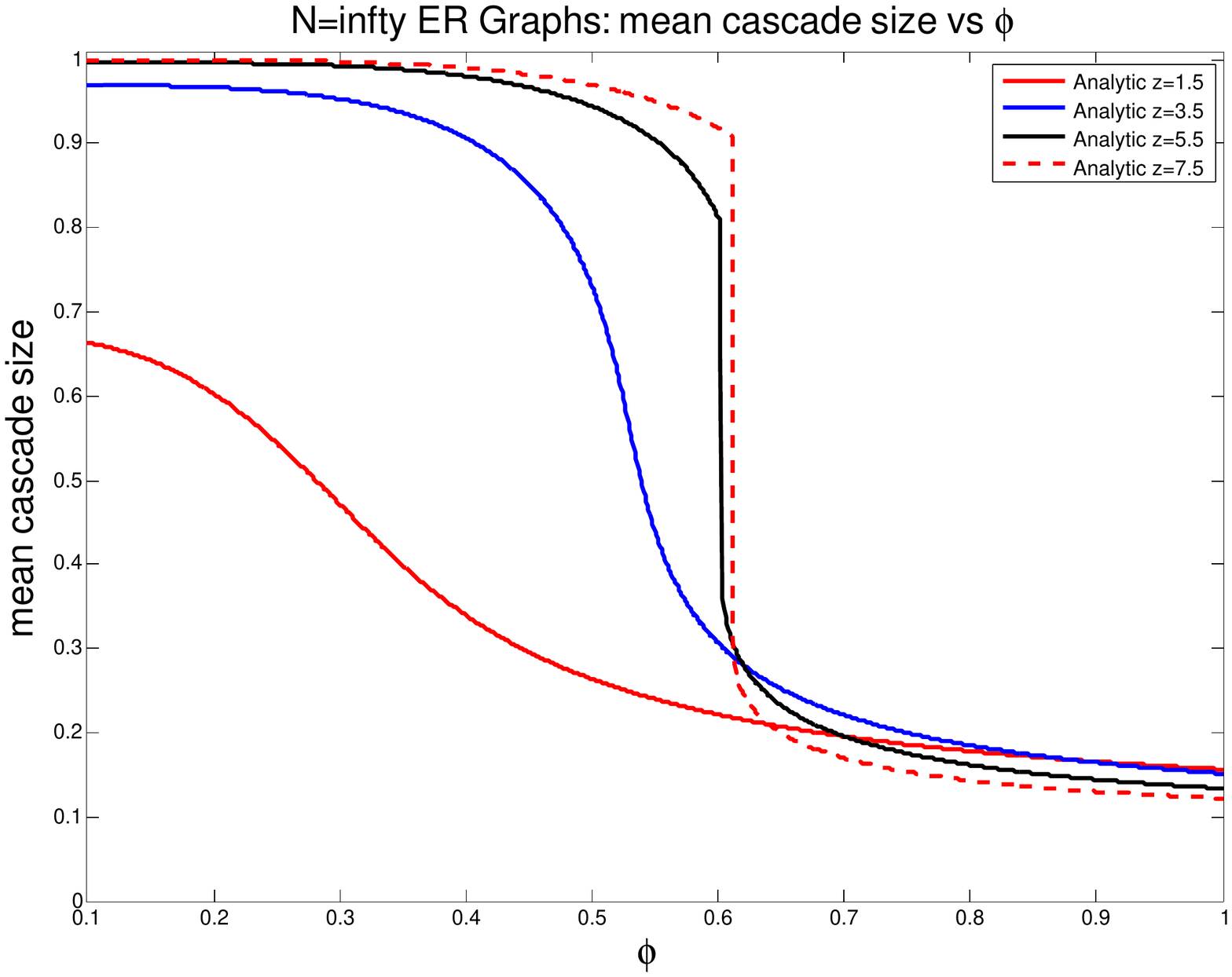}
\caption{Experiment 1C: This figure shows how the mean cascade size depends on both the connectivity parameter $z$ and the threshold parameter $\phi$, when the weights are lognormal with mean and standard deviation $(1,0.5)$.} \label{fig1C}
\end{figure}

In Experiment 1C, we observed the dependence of the average cascade size $\bar A^{(\infty)}$ on $z$ and $\phi$ when the edge weights were lognormal with mean and standard deviation $(1,0.5)$. We can see in  Figure \ref{fig1C}  that a jump discontinuity in $\bar A^{(\infty)}$ arises as a function of $\phi$, as $z$ increases above $3.5$. In the next section we will  observe that this discontinuity is strongly associated with discrepancies between the LTIA analytics and the true values for finite real world networks.

\subsection{Experiment 2: Finite Erd\"os-Renyi Graphs}

The aim of Experiment 2 was to investigate how the analytical formulas of Theorem \ref{thm2}, proved under the LTIA assumption, compare to the results of Monte Carlo simulations, on specific real world networks. In Experiment 2A we considered twelve Erd\"os-Renyi skeleton graphs, of three sizes $N=10,100, 1000$ and four different values of the connectivity parameter $z=1.5, 3.5, 5.5, 7.5$. In each case, the default seed was a random set of $N/10$ nodes, and the weights were log-normally distributed with mean and standard deviation $(1,0.5)$. The threshold distribution for a node of degree $k$ was log-normal with mean and standard deviation $(\phi k, \phi k/2)$ for variable $\phi$. 

\begin{figure}[ht]
\centering\vspace{-.8in} 
\includegraphics[scale=.56]%
{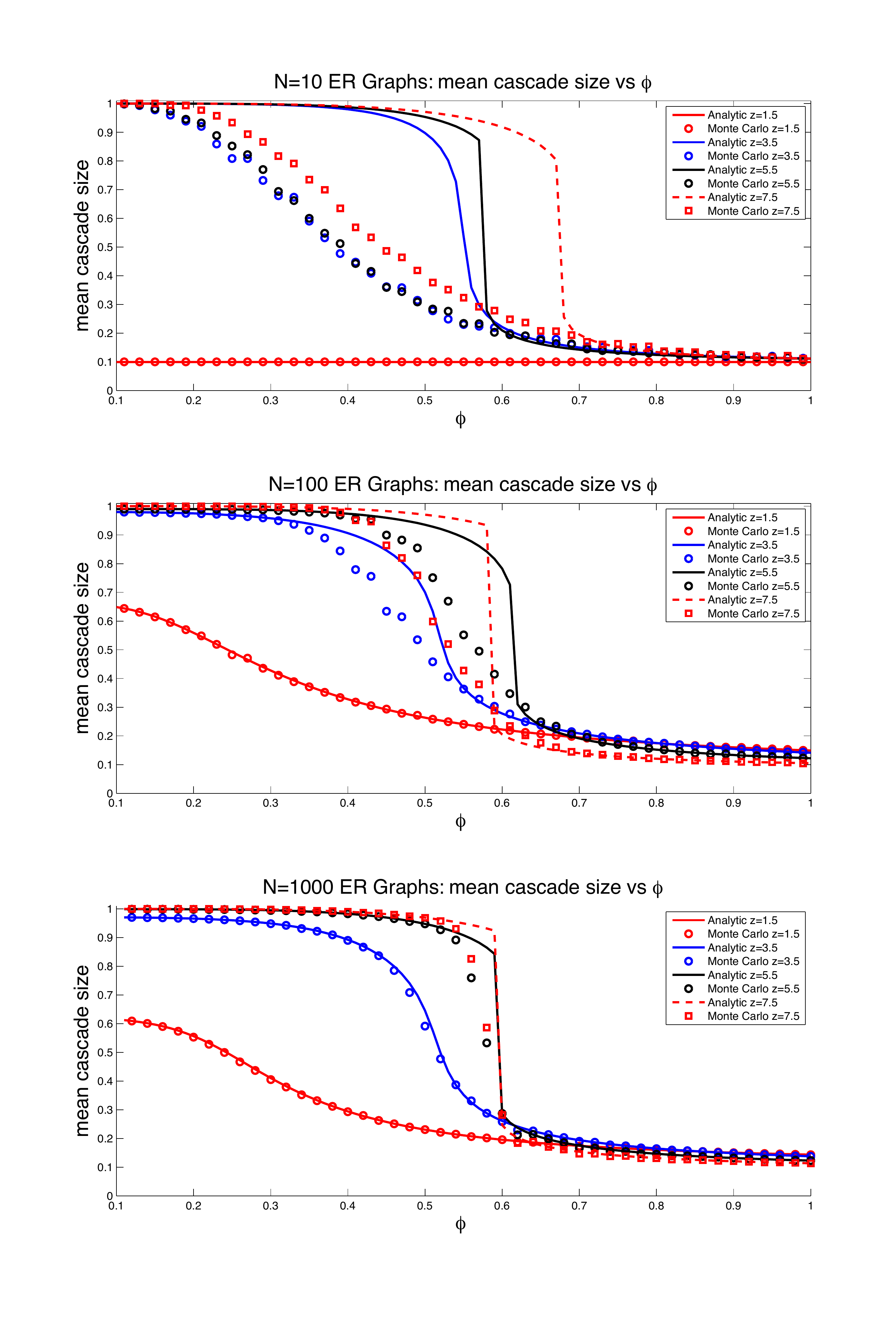}\vspace{-.8in} 
\caption{ Experiment 2A: These three figures shows how the size of global cascades on finite Erd\"os-Renyi skeleton graphs is influenced by lognormally distributed edge weights, with means consistent with the cases of Figure \ref{fig1A}. The figures from top to bottom show the results for skeleton graphs of size $10, 100, $ and $1000$ respectively; the four curves correspond to different values of the connectivity parameter $z$. In each case, both LTIA and Monte Carlo values are plotted as functions of the threshold parameter $\phi$.}\label{fig2A}
\end{figure}

In each of the three subplots of Figure \ref{fig2A}, the four curves show the analytical values of mean cascade size a functions of the threshold parameter $\phi$, for the four different values of $z$. The three subplots are for the three different values of $N$.  On each figure, the circles and squares show the corresponding Monte Carlo values computed with $10000$ realizations.

The Monte Carlo values shown on these figures give a very clear demonstration of how the $N\to\infty$ limit as shown in Figure \ref{fig1C} is approached through finite Erd\"os-Renyi graphs. One sees little discrepancy when $N=1000$, while for smaller $N$ the discontinuities in $\phi$ are smoothed out over a wider range of $\phi$. When one compares analytic results derived from Theorem \ref{thm2} to the Monte Carlo values, one finds discrepancies over a wide range of $\phi$ when $N=10$. However, for $N=1000$, the range over which a significant discrepancy occurs has narrowed considerably, and is confined to the range where the cascade size varies rapidly with $\phi$. Further numerical experimentation shows that this type of behaviour is quite typical.


\begin{figure}[ht]
\centering\vspace{-.8in} 
\includegraphics[scale=.6]%
{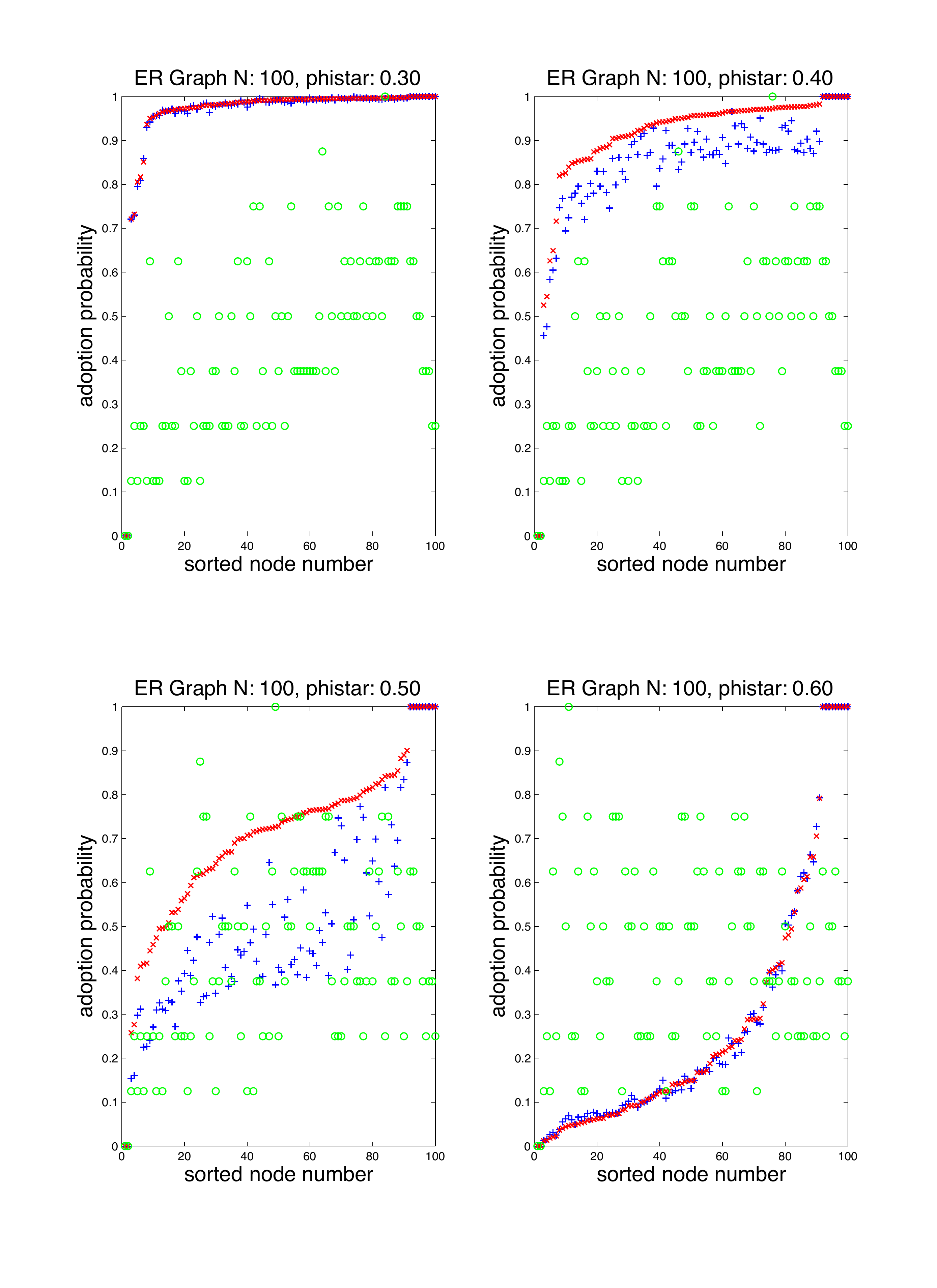}\vspace{-.6in} 
\caption{ Experiment 2B: These four panels show results for a single Erd\"os-Renyi skeleton graph with $N=100$ and $z=3.5$ and four different threshold parameter values $\phi=0.3, 0.4, 0.5, 0.6$. The panels plot the node-by-node $A_v^{(\infty)}$ probabilities as computed by both analytic (red crosses) and Monte Carlo methods (blue pluses), in increasing order of their analytic values. The green circles represent the node degree as a fraction of the maximum node degree.}\label{fig2B}
\end{figure}

Experiment 2B tested whether the LTIA analytic formula can accurately determine adoption probabilities on a node-by-node basis. The four panels of Figure \ref{fig2B} show results for a single Erd\"os-Renyi skeleton graph with $N=100$ and $z=3.5$ and four different threshold parameter values $\phi=0.3, 0.4, 0.5, 0.6$. The panels plot the node-by-node $A_v^{(\infty)}$ probabilities as computed by both analytic (red crosses) and Monte Carlo methods (blue pluses), in increasing order of their analytic values.  Taken together, they back up our general conclusion that the LTIA analytics show reasonable accuracy on a node-by-node basis, and that accuracy improves as one moves away from  medium cascade regions. The green circles represent the node degree as a fraction of the maximum node degree: one notices that there is very little indication that node degree is significantly correlated with adoption probability. 

\begin{remark}
We also ran consistency checks to verify that the Monte Carlo estimated cascades were unbiased estimators of the LTIA analytic cascades for a wide range of finite tree graphs. We created graphs similar to Figure \ref{fig2B} that in all cases showed the almost perfect agreement one expects for tree graphs. These results are not shown here.
\end{remark}

\section{Concluding Remarks}

The main result of this paper, the Cascade Theorem \ref{thm1} for models with random edge weights, is not a straightforward extension of original Watts Model. Unlike earlier versions of the theory, the inductive derivation of Theorem \ref{thm1} requires a subtle decoupling of the dependence of edge and threshold random variables. The key to this decoupling is the introduction of both the ``locally tree-like assumption'' (LTIA) and the ``without regarding'' (WOR) condition. We believe that this particular argument is new in the literature. For simplicity we have confined the discussion to information cascades: it is clear that the ideas extend to the more complex models of cascades in financial systems. 

We have shown both in infinite networks and in finite real world networks that the analytics of Theorems \ref{thm1} and \ref{thm2} give correct answers where appropriate, and are a meaningful approximation where the LTIA assumption fails to hold. We have used these results to look at the effects random edge weights have on information cascades, and  have found evidence of interesting transitions and ``tipping points'' deserving of further study. These theorems give us the possibility to explore the totality of models of this general type with efficient computer algorithms. Although our primary motivations come from the models of \cite{NieYanYorAle07} and \cite{GaiKapa10} for systemic risk in financial networks, it is possible that our results are applicable to a wide range of network phenomena.

\section{Acknowledgements}
This work was funded by awards from  Science Foundation Ireland (11/PI/1026 and MACSI
06/MI/005), from the FET-Proactive project PLEXMATH (grant no. 317614) funded by the European Commission, and from the Natural Sciences and Engineering Research Council of Canada. We acknowledge the SFI/HEA Irish Centre for High-End Computing (ICHEC) for the provision of computational facilities. We are indebted to our summer research intern, Lionel Cassier of \'Ecole Polytechnique France, who developed much of the preliminary numerical code that underlies the  graphics in this work. 

\bibliographystyle{abbrvnat}


\begin{thebibliography}{13}
\providecommand{\natexlab}[1]{#1}
\providecommand{\url}[1]{\texttt{#1}}
\expandafter\ifx\csname urlstyle\endcsname\relax
  \providecommand{\doi}[1]{doi: #1}\else
  \providecommand{\doi}{doi: \begingroup \urlstyle{rm}\Url}\fi
  




\bibitem[Amini et~al.(2012)Amini, Cont, and Minca]{AminContMinc12}
H.~Amini, R.~Cont, and A.~Minca.
\newblock {Stress Testing the Resilience of Financial Networks}.
\newblock \emph{Int. J. Theor. App. Fin.},
  15:\penalty0 1--20, 2012.

\bibitem[Bogu\~n\'a and Serrano(2005)]{BoguSerr05}
M.~Bogu\~n\'a and M.~A. Serrano.
\newblock Generalized percolation in random directed networks.
\newblock \emph{Phys. Rev. E}, 72:\penalty0 016106, 2005.

\bibitem[Bollob\~as(1980)]{Bollobas80}
B.~Bollob\~as.
\newblock A probabilistic proof of an asymptotic formula for the number of
  labelled regular graphs.
\newblock \emph{Eur. J. Comb.}, 1:\penalty0 311, 1980.

  \bibitem[Dodds and Payne(2009)]{DoddPayn09}
P.~S. Dodds and J.~L. Payne.
\newblock Analysis of a threshold model of social contagion on
  degree-correlated networks.
\newblock \emph{Phys. Rev. E}, 79:\penalty0 066115, 2009.

\bibitem[Dodds et~al.(2012)Dodds, Harris, and Danforth]{DodHarDan12}
P.~S. Dodds, K.~D. Harris, and C.~M. Danforth.
\newblock Limited imitation contagion on random networks: Chaos, universality,
  and unpredictability.
  \newblock arXiv: 1208.0255, 
2012.

\bibitem[Dodds et~al.(2012)]{DodHarPay11}
P.~S. Dodds, K.~D. Harris, and J.~L. Payne.
\newblock Direct, physically motivated derivation of the contagion condition for spreading processes on generalized random networks.
\newblock \emph{Phys. Rev. E}, 83:\penalty0 056122, 2011.

\bibitem[Easley and Kleinberg(2010)]{EaslKlei10}
D.~Easley and J.~Kleinberg.
\newblock \emph{Networks, Crowds and Markets}.
\newblock Oxford University Press, Oxford, 2010.

\bibitem[Eisenberg and Noe(2001)]{EiseNoe01}
L.~Eisenberg and T.~H. Noe.
\newblock Systemic risk in financial systems.
\newblock \emph{Management Sci.}, 47\penalty0 (2):\penalty0 236--249, 2001.

\bibitem[Erd\"os and R\'enyi(1959)]{ErdoReny59}
P.~Erd\"os and A.~R\'enyi.
\newblock On random graphs.
\newblock \emph{I. Publ. Math. Debrecen}, 6:\penalty0 290--297, 1959.

\bibitem[Erd\"os and R\'enyi(1960)]{ErdoReny60}
P.~Erd\"os and A.~R\'enyi.
\newblock On the evolution of random graphs.
\newblock \emph{Publ. Math. Inst. Hung. Acad. Sci.}, 5:\penalty0 17--61, 1960.

\bibitem[Gai and Kapadia(2010)]{GaiKapa10}
P.~Gai and S.~Kapadia.
\newblock Contagion in financial networks.
\newblock \emph{Proc. Roy. Soc. A}, 466\penalty0
  (2120):\penalty0 2401--2423, 2010.

\bibitem[Gleeson and Melnik(2009)]{GleeMeln09}
J.~P. Gleeson and S.~Melnik.
\newblock Analytical results for bond percolation and $k$-core sizes on
  clustered networks.
\newblock \emph{Phys. Rev. E}, 80:\penalty0 046121--046133, 2009.


\bibitem[Granovetter(1978)]{Granovetter78}
M.~Granovetter.
\newblock {Threshold Models of Collective Behavior}.
\newblock \emph{Am. J. Soc.}, 83:\penalty0 1420-1443, 1978.


\bibitem[Hurd and Gleeson(2011)]{HurdGlee11}
T.~R. Hurd and J.~P. Gleeson.
\newblock A framework for analyzing contagion in banking networks.
\newblock arXiv:1110.4312 [q-fin.GN], October 2011.

\bibitem[Melnik et~al.(2011)Melnik, Hackett, Porter, Mucha, and
  Gleeson]{MelHacMasMucGle11}
S.~Melnik, A.~Hackett, M.~A. Porter, P.~J. Mucha, and J.~P. Gleeson.
\newblock The unreasonable effectiveness of tree-based theory for networks with
  clustering.
\newblock \emph{Phys. Rev. E}, 83:\penalty0 036112--036123, 2011.

\bibitem[Newman(2010)]{Newman10}
M.~E.~J. Newman.
\newblock \emph{Networks: An Introduction}.
\newblock Oxford University Press, 2010.

\bibitem[Nier et~al.(2007)Nier, Yang, Yorulmazer, and Alentorn]{NieYanYorAle07}
E.~Nier, J.~Yang, T.~Yorulmazer, and A.~Alentorn.
\newblock {Network Models and Financial Stability}.
\newblock \emph{J. Econ. Dyn. Control}, 31:\penalty0 2033--2060, 2007.


\bibitem[Schelling(1973)]{Schelling73}
T.~C. Schelling.
\newblock {Hockey Helmets, Concealed Weapons, and Daylight Saving: A Study of Binary Choices With Externalities}.
\newblock \emph{J. Conflict Resolution}, 17:\penalty0 381-428, 1973.

\bibitem[Traud et~al.(2011)Traud, Kelsic, Mucha, and Porter]{TraKelMucPor11}
A.~L. Traud, E.~D. Kelsic, P.~J. Mucha, and M.~A. Porter.
\newblock Comparing community structure to characteristics in online collegiate
  social networks.
\newblock \emph{SIAM Rev.}, 53:\penalty0 526--543, 2011.

\bibitem[Watts and Dodds(2007)]{WattDodd07}
D.~J. Watts and P.~S. Dodds.
\newblock Influentials, networks, and public opinion formation.
\newblock \emph{J. Consumer Res.}, 34:\penalty0 441--458, 2007.

\bibitem[Watts(2002)]{Watts02}
D.~J. Watts.
\newblock A simple model of global cascades on random networks.
\newblock \emph{PNAS}, 99\penalty0 (9):\penalty0 5766--5771, 2002.

\end{thebibliography}

\end{document}